# Tunable itinerant spin dynamics with polar molecules

Jun-Ru Li, Kyle Matsuda, Calder Miller, Annette N. Carroll, William G. Tobias, Jacob S. Higgins, Jun Ye

[1]JILA, National Institute of Standards and Technology and Department of Physics, University of Colorado, Boulder, CO 80309.

*Corresponding authors. Emails: junru.li@colorado.edu, ye@jila.colorado.edu

**Strongly interacting spins underlie many intriguing phenomena and applications[1-4] ranging from magnetism to quantum information processing. Interacting spins combined with motion display exotic spin transport phenomena, such as superfluidity arising from pairing of spins induced by spin attraction[5,6]. To understand these complex phenomena, an interacting spin system with high controllability is desired. Quantum spin dynamics have been studied on different platforms with varying capabilities[7-13]. Here we demonstrate tunable itinerant spin dynamics enabled by dipolar interactions using a gas of potassium-rubidium molecules confined to two-dimensional planes, where a spin-1/2 system is encoded into the molecular rotational levels. The dipolar interaction gives rise to a shift of the rotational transition frequency and a collision-limited Ramsey contrast decay that emerges from the coupled spin and motion. Both the Ising and spin exchange interactions are precisely tuned by varying the strength and orientation of an electric field, as well as the internal molecular state. This full tunability enables both static and dynamical control of the spin Hamiltonian, allowing reversal of the coherent spin dynamics. Our work establishes an interacting spin platform that allows for exploration of many-body spin dynamics and spin-motion physics utilizing the strong, tunable dipolar interaction.**

Ultracold polar molecules, with (pseudo-)spins encoded in the molecules' internal states, provide a unique platform for investigating many-body spin dynamics with high tunability[13-20]. The effective spin coupling is realized via long-range, anisotropic dipolar interactions mediated by the molecule's electric dipole moment, which is tunable with external electric fields and microwaves. Pioneering proposals[14-16, 19] and experiment[13] used polar molecules localized in optical lattices to realize lattice-spin models. Even with modest lattice fillings, the strong dipolar interaction allowed the observation of spin-exchange dynamics[13]. Recently, spin correlations have been directly imaged under a microscope[12]. The strong tunability of this lattice-spin system enables the realization of a variety of spin Hamiltonians displaying exotic correlations and phase transitions[16,17,19-21].

In parallel, molecules loaded in a two-dimensional (2D) layer are allowed to move freely within the plane, and dipolar interactions control the collisional process. The repulsion from the dipolar interaction enhances elastic collisions while suppressing reactive losses, allowing evaporative cooling to quantum degeneracy[22]. The engineered dipolar interactions create a strong repulsive energy barrier that allows control of the molecular reaction with only a small change of the electric field[23,24] or microwaves[25,26]. In similar systems with magnetic atoms, magnetic dipolar interactions, albeit relatively weak, yield a variety of quantum phases and dynamics[11].

Combining coherent spin interactions and motional physics in a single platform is expected to give rise to rich dynamics. In this article, we report the realization of an itinerant spin system that evolves from a fully controlled spin Hamiltonian to a more complex spin-motion coupled system. Isolated 2D layers of dipolar molecules are addressed by a controlled external electric field ($E$) to realize both Ising and exchange interactions with high tunability (Fig. 1**a**). At short evolution times, the system's dynamics are governed by dipolar interactions in the spin degrees of freedom, realizing a spin Hamiltonian that can generate highly entangled spin states[27]. At longer times where collisions mediated by the molecular thermal motion and dipolar interaction manifest, we observe the onset of spin decoherence that is not reversible by engineered echo pulse sequences. Our work establishes a tunable dipolar system where the spin and motional dynamics are coupled by strong dipolar interactions.

We create gases of fermionic $^{40}$K$^{87}$Rb (KRb) loaded into 2D harmonic traps formed by a one-dimensional optical lattice[22,28] (see Methods). The spin-1/2 degree of freedom is mapped onto molecular rotations as $|\downarrow\rangle = |0, 0\rangle$ and $|\uparrow_1\rangle = |1, 0\rangle$ or $|\uparrow_2\rangle = |1, -1\rangle$ (Fig.1b). Here, $|N, m_N\rangle$ denotes a rotational state with $N$ being the field-dressed rotational quantum number and $m_N$ its projection on the quantization axis set by $\boldsymbol{E}$ (or the magnetic field along $\hat{\boldsymbol{y}}$ at $|\boldsymbol{E}| = 0$). Spin states are prepared using a microwave field to drive the molecules' rotational transitions. The magnitude and orientation of $\boldsymbol{E}$ and the molecular spin states are used to tune the dipolar interaction, and consequently, the spin and collision dynamics.

At short times prior to the onset of collisions, the system evolves primarily through the spin-1/2 degree of freedom, as molecular motion is effectively frozen. Dipolar interactions couple individual molecular spins $\boldsymbol{s}_i = \{s_i^X, s_i^Y, s_i^Z\}$ occupying the harmonic oscillator mode $i$, leading to spin dynamics described by an XXZ Heisenberg Hamiltonian[27]: $H = \frac{1}{2}\sum_{ij}\left[J_{ij}^Z s_i^Z s_j^Z + J_{ij}^\perp (s_i^X s_j^X + s_i^Y s_j^Y)\right] + \sum_i s_i^Z h_i^Z$. Here, $h_i^Z$ is the effective field. $J_{ij}^Z$ and $J_{ij}^\perp$ are the Ising and the exchange interactions arising from the $\boldsymbol{E}$-induced dipole moments $d_\downarrow$ and $d_\uparrow$ and the transition dipole moment $d_{\downarrow\uparrow}$, respectively.

In contrast to lattice spins[13], spins delocalized in the 2D plane have large and nearly uniform spatial wavefunction overlap between any pairs of molecules. As a result, $J_{ij}^\perp, J_{ij}^Z$, and $h_i^Z$ have only weak dependence on $\{i, j\}$, leading to an effective description of dynamics in terms of the ensemble averaged values $J_\perp, J_Z, h_Z$ and collective spin[27] $S_\gamma = \sum_i s_i^\gamma$ ($\gamma = X, Y, Z$):

$$H = J_\perp S^2 + \chi_o S_Z^2 + h_Z S_Z . \qquad (1)$$

$J_\perp$ enforces an energy cost to change $|S|$ and thus protects the collective spin from decoherence. $h_Z$ represents an effective magnetic field. The strength of spin-spin interaction is characterized by $\chi_o = J_Z - J_\perp$, which is the difference between the average Ising and exchange interaction. This term produces a one-axis twisting that can generate highly entangled spin states[27,29,30]. All these terms are tunable with $\boldsymbol{E}$ or the choice of spin states. In particular, both the magnitude and sign of the interaction parameter $\chi_o \propto (3\cos^2\alpha - 1)[(d_\downarrow - d_\uparrow)^2 - 2d_{\downarrow\uparrow}d_{\uparrow\downarrow}]$ (see Methods) can be varied by changing the angle $\alpha$ between $\boldsymbol{E}$ and $\hat{\boldsymbol{y}}$ (Fig. 1a) or $|\boldsymbol{E}|$, as exemplified by the spin manifold $\{|\downarrow\rangle,|\uparrow_1\rangle\}$ where $d_{\downarrow\uparrow} = d_{\uparrow\downarrow} = \langle\downarrow |d| \uparrow_1\rangle$ (Fig. 1c).

At long times, dipolar interactions lead to elastic collisions that change the motional modes of the molecules at a rate depending on the molecular spins and coherence. In the motional degree of freedom, these mode-changing collisions lead to dipolar thermalization[22]. In the spin degree of freedom, they cause decoherence of the collective spin at the collision rate because each of the collisional events introduces *random* spin rotations on the two colliding spins. This decoherence cannot be reversed due to scrambled spin-motion coupling.

Our system exhibits unique dynamics arising from the molecular spin and motion intertwined by dipolar interactions: Delocalized molecules facilitate the couplings between molecular spins, and spin dynamics affect the molecular collisions through spin coherence. Using dynamical decoupling techniques, we characterize both effects with Ramsey spectroscopy. The former manifests as a frequency shift on the spin transition, the latter as an interaction dependent fringe contrast decay.

We first characterize the spin dynamics by measuring the frequency shift arising from $\chi_o$. At the mean-field level, $\chi_o S_Z^2 \approx 2\chi_o S_Z \langle S_Z \rangle$ produces a shift to the spin transition frequency that is proportional to $\langle S_Z \rangle$, representing a population imbalance of $(N_\uparrow - N_\downarrow)/2$ in a coherent superposition of the two spin states. Here, $N_\uparrow$ and $N_\downarrow$ are the numbers of molecules in each spin state. For a Ramsey experiment with an interrogation time $T$ and an initial pulse area $\theta$ that prepares the system with $\langle S_Z \rangle = -[(N_\uparrow + N_\downarrow)/2]\cos\theta$, this leads to a dipolar interaction-induced fringe phase shift

$$\Delta\phi/2\pi = -n\chi T \cos\theta. \qquad (2)$$

Here, $\chi = (\chi_o/h)[(N_\uparrow + N_\downarrow)/n]$, where $n$ is the average 2D molecular density and $h$ is the Planck constant. The strength of the mean-field interaction $U_\chi$ is thus defined as $U_\chi = n\chi$.

For a typical density $n = 1 \times 10^7$ cm$^{-2}$, the frequency shift ($\Delta\phi/2\pi T$) is expected to be about 100 Hz. Measuring such shifts requires an appropriately long spin coherence time. However, the experimentally achieved coherence time $T_2^*$ between molecular rotations is limited by single particle dephasing mechanisms[28,31]. Grey circles in Fig. 2**b** show decay of the Ramsey contrast for $|E| = 0$ under typical trapping conditions. A Gaussian fit gives $T_2^*$ of only 0.24(1) ms. Extending $T_2^*$ to milliseconds has been demonstrated by orienting the bias electric field to the *magic angle*[28,31-33]. However, achieving long $T_2^*$ in an arbitrary configuration of $E$, which is required for tuning of the dipolar interaction, has been challenging.

We perform dynamical decoupling with an XY8 multi-pulse sequence[34], consisting of eight Rabi-$\pi$ rotations along $\hat{X}$ and $\hat{Y}$ spaced by time $\tau$ (Fig. 2a), to suppress single particle dephasing. Multiple XY8 sequences can be concatenated (denoted XY8×$M$) to achieve longer $T$ (see Methods). Implementing XY8×3 leads to a factor of 71(5) improvement in $T_2^*$ from 0.24(1) ms to 17(1) ms for the same field and trapping condition (orange squares in Fig. 2b), sufficient for studying the dipolar density shift as a function of $|E|$ and $\alpha$.

To measure the frequency shift and thus $\chi$, a microwave pulse with an area $\theta$ first prepares KRb in a superposition state $\cos(\theta/2)|\downarrow\rangle + \sin(\theta/2)|\uparrow_1\rangle$. After a sequence of XY8 with $T$ = 1.2 ms (Fig. 2a), the accumulated phase is mapped onto population by a second Ramsey pulse with an area of $\pi$-$\theta$ and a variable phase $\phi$. The fraction in $|\uparrow_1\rangle$ as a function of $\phi$ are subsequently measured by imaging molecules in both states[28], from which we obtain Ramsey fringes like those shown in Fig. 2c for $|E|$ = 0. By fitting the fringes to a sinusoidal function, we extract the corresponding phase shifts $\Delta\phi$. At a fixed $n$, we measure $\Delta\phi$ at several different values of $\theta$ and obtain $U_\chi$ as the slope of the linear fit to $\Delta\phi$ versus $\cos\theta$ (solid black and solid grey lines in Fig. 2d). We repeat the measurement of $U_\chi$ at several values of $n$ to measure $\chi$. The results for $|E|$ = 0 are shown with grey circles in Fig. 2e. The measured $U_\chi$ as a function of $n$ is well captured by a linear fit, consistent with the mean-field model. From the slope of the fit, we extract $\chi_1$ = -4.9(3) × $10^{-6}$ Hz/cm$^{-2}$ at $|E|$ = 0 between $|\downarrow\rangle$ and $|\uparrow_1\rangle$.

Leveraging a configurable $|E|$ that mixes the molecular rotational states allows us to vary the relative strength between the Ising and exchange interactions in the spin dynamics. We demonstrate this tunability by measuring $\chi$ over a wide range of $|E|$ at $\alpha$ = 0° for the manifold $\{|\downarrow\rangle, |\uparrow_1\rangle\}$. The results, shown in Fig. 2f, are well captured by scaling $\chi \propto [(d_\downarrow - d_\uparrow)^2 - 2d_{\downarrow\uparrow}d_{\uparrow\downarrow}]$ (black solid line) to fit the experimental data. At low $|E|$, dipolar exchange dominates and $\chi < 0$. As $|E|$ increases, the increasing strength of the Ising interactions from the induced dipoles competes with the spin exchange eventually leading to $\chi > 0$ (Fig. 1e) at high $|E|$. Tuning $|E|$ to approximately 7 kV/cm where $\chi = 0$ would realize the Heisenberg XXX model. In future studies, such tunability will allow exploration of qualitatively different spin dynamics hosted by a spin-1/2 system[16].

In 2D, the average dipolar interaction can be controlled by orienting the dipole moments relative to the 2D plane, as a direct consequence of the anisotropy of the dipolar interaction. We measure $\chi$ between $|\downarrow\rangle$ and $|\uparrow_1\rangle$ as a function of $\alpha$ at $|\boldsymbol{E}| = 1.02$ kV/cm. By rotating the dipoles, $\chi$ is varied continuously from negative to positive (Fig. 2**g**). At $\alpha = 0°$, molecules interact repulsively as the dipole moments are aligned perpendicular to the 2D plane, leading to $\chi < 0$. At $\alpha = 90°$, dipoles are aligned within the plane and the average interaction is attractive, leading to $\chi > 0$. Assuming symmetric transverse trapping, $\chi \propto -(3\cos^2\alpha - 1)$. Notably, this tunability of $\chi$ by rotating $\boldsymbol{E}$ arises from a common geometric factor on the Ising and exchange interactions, in contrast to tuning with $|\boldsymbol{E}|$, which controls the relative strength of the two interactions.

The choice of the internal states used for the spin-1/2 system offers additional control over the interaction parameters. The green squares in Fig. 2**e** show a measurement of $U_\chi$ between $|\downarrow\rangle$ and $|\uparrow_2\rangle = |1, -1\rangle$ at $|\boldsymbol{E}| = 0$. Using $|\uparrow_2\rangle$ instead of $|\uparrow_1\rangle$ changes the magnitude of $\chi$ and reverses its sign at $|\boldsymbol{E}| = 0$, as a result of the different transition dipole moments in $\{|\downarrow\rangle, |\uparrow_2\rangle\}$ where $d_{\uparrow\downarrow} = -d_{\downarrow\uparrow} = \langle\downarrow|d|\uparrow_1\rangle/\sqrt{2}$. We measure $\chi$ between $|\downarrow\rangle$ and $|\uparrow_2\rangle$ to be $\chi_2 = 2.3(8)\times10^{-6}$ Hz/cm$^{-2}$, consistent with $-1/2$ of the interaction between $|\downarrow\rangle$ and $|\uparrow_1\rangle$ at this field.

The ability to manipulate dipolar interactions through the internal state allows us to dynamically change the interaction and the evolution of the many-body state. In contrast to tuning with the dc electric field, internal state control of the interaction can be achieved rapidly and coherently with microwave pulses. Such a capability is essential for applications such as dynamic Hamiltonian engineering[35], high fidelity quantum gate operations[36], and studying quantum quenched dynamics with dipolar spin systems[37,38].

By implementing a dynamical control pulse sequence, we demonstrate reversal of the coherent many-body spin dynamics. In particular, we coherently swap the excited state from $|\uparrow_2\rangle$ to $|\uparrow_1\rangle$ in the middle of the Ramsey evolution, which instantaneously changes the interaction parameter $\chi$ by a factor of $-2$, and consequently reverses the associated phase progress in a coherent evolution.

Our sequence consists of two phase accumulation stages (Fig. 3**a**). After preparation in a superposition of $|\downarrow\rangle, |\uparrow_2\rangle$, molecules evolve with $\chi_2$ (stage I) until a composite microwave pulse $R$ (Inset of Fig. 3**a**) coherently transfers the excited state population from $|\uparrow_2\rangle$ to $|\uparrow_1\rangle$ while

maintaining the phase relative to $|\downarrow\rangle$. Molecules then continue to evolve in $\{|\downarrow\rangle, |\uparrow_1\rangle\}$ (stage II) with $\chi_1 = -2\chi_2$ before being detected after the final Ramsey pulse. The XY8 decoupling sequence is used during both stages. The duration of $R$ is 70 μs, which is short compared to the interaction timescales. We then extract the phase shift $\Delta\Phi = \Delta\phi(\theta = \pi/4) - \Delta\phi(\theta = 3\pi/4)$ as a function of $T$, plotted in Fig. 3**b**. (see Methods)

We observe a sign change in the rate of phase accumulation $d(\Delta\Phi)/dT$ at $t_s = 1.2$ ms when $R$ is implemented, indicating a reversal of the mean-field interaction sign. The data is well described by a piecewise linear fit with the ratio of $d(\Delta\Phi)/dT$ before and after the reversal constrained to be $-2$, consistent with the measurement in Fig. 2**e**. At $T \approx 1.6$ ms, $\Delta\Phi$ returns to the initial value, completing the phase reversal.

Our protocol realizes a *complete* reversal of the spin Hamiltonian at $|E| = 0$. Such a capability is essential for studying out-of-time-ordered correlators which are used to understand dynamics of interacting quantum many-body systems such as quantum information scrambling[39-42]. Similar many-body echo processes also allow applications such as robust Heisenberg-limited phase sensitivity without single-particle-resolved state detection[27,43-45].

At long evolution times, motional and spin dynamics are coupled by dipolar interactions, resulting in dynamical evolution beyond the spin model. Mode-changing dipolar collisions lead to loss of coherence, manifesting as an exponential decay of the Ramsey contrast, $\exp(-\Gamma t)$. $\Gamma$ is proportional to the rate of elastic collisions $n\sigma v_T$, where $v_T$ is the thermal velocity and $\sigma$ is the cross section for dipolar elastic collisions. For spin-polarized fermionic molecules with dipole moment $d$ induced by $E$, previous theories[46,47] and experiments[22,24] have shown $\sigma \propto d^4$. By comparison, two molecules in a coherent superposition carry oscillating dipoles with amplitudes of $d_{\downarrow\uparrow}$ in the lab frame, leading to dipolar collisions even at low $|E|$ where induced dipole moments are small[26,48]. When the oscillations are fully coherent, the molecules collide with maximum effective dipole moment $d_{\downarrow\uparrow}/\sqrt{2}$ and $\sigma$ (Ref. 48). These collisions couple the spin and motion, resulting in spin decoherence that cannot be removed using multi-pulse sequences, in contrast to interaction-induced dephasing in a lattice[13].

We study the effect of dipolar collisions by measuring contrast decay as a function of $n$. To investigate collisional decoherence, the single particle dephasing rate must be below the collisional rate, which was achieved by increasing the number of decoupling pulses.

We observe the characteristic exponential decay of the contrast for a range of $n$ (Fig. 4**a**). The extracted $\Gamma$ increases with density (Fig. 4**b**), indicating collision-limited coherence time. At $n = 1.1(1) \times 10^7$ cm$^{-2}$ and $|E| = 0$, we measure $\Gamma = 130(5)$ s$^{-1}$, consistent with dipolar elastic collision rates measured in Ref. 22 with $d$ being $d_{\downarrow\uparrow}/\sqrt{2}$ (see Methods). This gives a collisional timescale shorter than the trap oscillation period of ~22 ms (see Methods). Rotating the electric field to $\alpha = 36°$ at $|E| = 1.02$ kV/cm reduces the strength of the dipolar interaction. We observe longer coherence times at the same density (purple squares in Fig. 4**b**), indicating reduced $\sigma$.

By extracting the average $\sigma = \Gamma/nv_T$ for each $E$, we found $\sigma(\alpha = 0°, |E| = 0)/\sigma(36°, 1.02$ kV/cm$) = 3.1(4)$. This variation of $\sigma$ is expected to arise from the changing dipolar cross section as a function of $\alpha$, similar to scattering with induced dipole moments[24,47]. In analogy to the relationship in atomic systems between the mean-field shift and elastic cross section, the variation of $\sigma$ could also be explained by a scaling $\sigma \propto \chi^2$. The measured ratio of $\sigma$ is consistent with $[\chi(0°, 0)/\chi(36°, 1.02$ kV/cm$)]^2 = 3.5(9)$ from Fig. 2**f**, 2**g**. These results suggest that dipolar effects dominate the collisional decoherence process.

The high controllability of our system allows tuning the motional dynamics relative to the coherent spin dynamics by changing parameters such as the strength of optical trapping[49], temperature, and dipolar interaction strength. These capabilities would allow the investigation of other many-body spin-motion effects such as unconventional paired superfluidity[50,51], spin-wave and spin transport physics[52,53], and dipole-mediated spin-orbital dynamics[54].

In conclusion, we have demonstrated tunable dynamics of the collective spin and motion in a 2D itinerant spin system formed from a gas of fermionic polar molecules. We have shown static and dynamical control over the spin Hamiltonian using external fields, microwaves, and internal states of the molecules. Our results establish a highly controllable spin system that can be used to investigate a broad range of many-body phenomena.

# Methods

**Molecular spin system and preparation**

We create ultracold gases of KRb in 2D with the procedure detailed in Ref. 22 and Ref. 28. In brief, degenerate gases of $^{40}$K and $^{87}$Rb are loaded into a stack of 2D harmonic traps formed by a one-dimensional optical lattice. Molecules in the ground rotational state $|\downarrow\rangle$ are subsequently produced by magneto-association and stimulated Raman adiabatic passage (STIRAP) directly at the electric field $E$. We typically produce 15×10$^3$ molecules at a temperature $T_0 \approx 450$ nK, occupying 19(1) layers with a peak number of about 1000 molecules in a single layer. The harmonic trapping frequencies within each layer for molecules in $|\downarrow\rangle$ at $|E| = 0$ are $(\omega_x, \omega_y, \omega_z) = 2\pi \times (45, 17000, 45)$ Hz. Since the harmonic level spacing $\hbar\omega_y$ along the tight confinement direction is larger than $k_B T_0$ and the interaction energy scale, our molecules predominantly remain in the lowest harmonic level, forming a quasi-2D system throughout the experiments. Here, $k_B$ is the Boltzmann constant.

Couplings between these 2D layers are weak compared to the intralayer interaction[28] due to the large spatial extension of the fermionic molecular gases in each layer that averages out the interlayer interaction. This weak interlayer interaction is expected to manifest as only a small correction to $\chi$ for the timescales considered in this work. We therefore approximate the system as a stack of isolated layers.

At $|E| = 0$ and typical trapping condition, the resonant transition frequency between $|\downarrow\rangle$ and $|\uparrow_1\rangle$ ($|\downarrow\rangle$ and $|\uparrow_2\rangle$) is measured to be around 2228.138 MHz (2227.742 MHz). Compared to our typical Rabi frequency of 100 kHz, the large difference between resonant frequencies allows us to resolve these states by changing the microwave frequency. The degeneracy of $|1,1\rangle$ and $|\uparrow_2\rangle = |1, -1\rangle$ is broken by a weak coupling of the nuclear and rotational degrees of freedom, resulting in an energy difference of about 100 kHz between the two states. To avoid off-resonant transfer to $|1,1\rangle$ when driving the transition between $|\downarrow\rangle$ and $|\uparrow_2\rangle$, we reduce the microwave Rabi frequency to about 60 kHz.

Besides rotational structure, molecules possess hyperfine states. In our experiment, KRb is prepared in $|\downarrow\rangle = |N = 0, m_N = 0, m_K = -4, m_{Rb} = 1/2\rangle$. Here, $m_K$ and $m_{Rb}$ are the projection of the

nuclear spins on the quantization axis. For states with $N = 1$ ($|\uparrow_1\rangle$ and $|\uparrow_2\rangle$), electric quadrupole interaction slightly mixes the hyperfine states and the rotational states. Since hyperfine changing transitions are weak and their energy spectrum is well resolved, we use states with the largest projection on $|1, 0, -4, 1/2\rangle$ and $|1, -1, -4, 1/2\rangle$ as $|\uparrow_1\rangle$ and $|\uparrow_2\rangle$, respectively. In this work, we avoid driving hyperfine-changing transitions, and thus neglect the hyperfine degree of freedom.

The density of the molecular gas is varied by adding an extra cleaning procedure after the molecules are produced and before the Ramsey interrogation. Specifically, a microwave pulse with an area of $\theta_c$ prepares molecules in a superposition of $|\downarrow\rangle$ and $|\uparrow_1\rangle$. The STIRAP beam that is resonant with $|\downarrow\rangle$ and the electronically-excited intermediate state of the molecules is switched on for 50 μs to eliminate the $|\downarrow\rangle$ component of the superposition. The remaining molecules are then flipped back to $|\downarrow\rangle$ with a π pulse. The density is reduced to $\sin^2(\theta_c/2)$ of the initial density. At certain values of $E$, the cleaning STIRAP light pulse can also off-resonantly excite molecules in $|\uparrow_1\rangle$. In that case, molecules in $|\uparrow_1\rangle$ are shelved in $|2,0\rangle$ during the optical blast using an additional microwave pulse, and no loss of $|\uparrow_1\rangle$ molecules is observed.

Since the size of the STIRAP beam used for cleaning is larger than the size of the sample, we do not observe significant temperature change from the process. This method therefore allows us to control the sample density without changing the temperature or the spatial distribution in the regime where the experiments are conducted.

**Timescales of the dynamics**

Multiple timescales are involved in the dynamics presented in this work: 1) interaction timescale set by $U_\chi$, 2) collision timescale set by $\Gamma$, and 3) single-particle motion timescale set by the period of harmonic trapping.

The spin model is valid in the regime $t < 1/\Gamma$ (short time limit). For $n = 1.1(1) \times 10^7$ cm$^{-2}$ (where the strongest collisional effects are explored in the current work), $1/\Gamma \approx 7.4$ ms. At this density, $U_\chi \approx 60$ Hz, giving a timescale of $1/U_\chi \approx 17$ ms. Since $U_\chi \propto \chi$ while $\sigma$, therefore $\Gamma$, scales with $\chi^2$, it is possible to tune the relative timescale of these two processes. Spin dynamics are favored for small $\chi$, while collisional dynamics dominate at large $\chi$.

The period of trap oscillation is ~22 ms. Therefore, for the highest density ($1/\Gamma \approx 7.4$ ms), molecules collide before they complete one trap oscillation (akin to the hydrodynamic regime). For the lowest collision rate we measured when we lowered the density or reduced the strength of dipolar interaction, molecules would undergo several oscillations before a collision occurs (collisionless regime). The relative rates of the two processes can be controlled by changing the strength of the optical trapping, temperature, density, or $\chi$.

The capability of tuning the system across different regimes highlights the controllability of our system.

**2D density calibration**

We use the following procedure to obtain the average 2D densities in the layers from the total molecule number $N_{tot}$. The molecules occupy multiple layers formed by an optical lattice. We first measure the molecule distribution using layer-resolved spectroscopy[28]. In brief, an electric field gradient along the longitudinal direction of the optical lattice shifts the transition frequency between $|0,0\rangle$ and $|1,0\rangle$ on each layer. When the differential shift for molecules in adjacent layers is large compared to the linewidth of the transition in each layer, molecules in different layers can be addressed, resolved, and measured spectroscopically using microwaves. A typical measurement (Extended data figure 1) reveals a Gaussian distribution with a full-width-half-maximum of 11(1) layers.

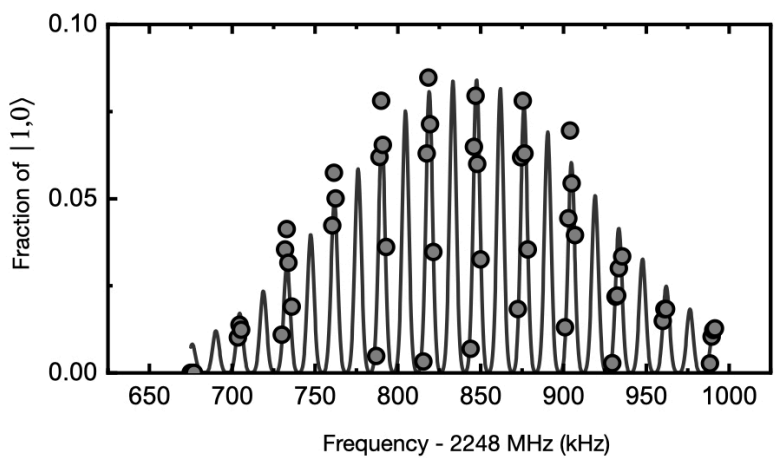

**Extended data figure 1. A typical molecular number distribution measured via layer-resolved spectroscopy.**

Grey circles are the experimental measurements. Black solid line is a fit to a summation of equally spaced Gaussian functions with a global Gaussian envelope. The width of each narrow Gaussian peak is assumed to be the same. The data is taken at $|E|$ = 1.02 kV/cm and $\alpha$ = 36° (magic condition) to reduce broadening of the single-layer transition linewidth due to differential polarizability.

For two-body processes considered in this work, such as the mean-field frequency shift and collisional decoherence, we calculate the average number of molecules per layer using an effective number of layers $L$, defined as $L = \frac{N^2}{\sum_k N_k^2}$, where $N_k$ is the number for molecules in the $k$-th layer. From the layer-resolved number distribution, we extract $L$ = 19(1) and use $N_{2D} = N_{tot}/L$ as the average number of molecules in a single layer. The average 2D density $n$ is calculated from $N_{2D}$ using temperatures measured with time-of-flight thermometry and the transverse trapping frequencies for $|\downarrow\rangle$ at $E$.

Our procedure for varying the density is not expected to change the number distribution across layers. We therefore use $L$ = 19 for all our measurements.

**Geometric factor of $\chi$**

The dipolar interaction between two dipoles has a geometric factor $(1-3\cos^2\Theta_{ij})$, where $\Theta_{ij}$ is the angle between the dipole moment and the vector between the two particles. When the molecular motion is confined to 2D, $\cos^2\Theta_{ij}$ can be decomposed into $\alpha$ and the azimuth angle $\beta$ as $\cos^2\Theta_{ij} = \sin^2\alpha \cos^2\beta$. This geometric factor can be expressed in terms of $\alpha$ by taking the average over the harmonic oscillator states in the 2D plane. In the mean-field regime and with approximately symmetric transverse trapping, this averaging reduces to an average over $\beta$, yielding $\langle\cos^2\beta\rangle = 1/2$ and thus $\langle 1-3\cos^2\Theta_{ij}\rangle = (3\cos^2\alpha -1)/2$.

**Measurement of Ramsey coherence time**

For the coherence time measurements, we measure the contrast at different Ramsey time $T$ with the number of XY8 echo pulses fixed. To extract the contrast at each $T$, we perform 8 to 16 measurements of the fraction of $|\uparrow_1\rangle$ with $\phi$ equally spaced between 0 and 360 degrees. The contrast and its standard deviation (SD) at time $T$ are extracted from the measured fractions via bootstrapping following the procedure described in Ref. 28.

**Dynamical decoupling and calibration of the pulse sequence**

We perform dynamical decoupling with an XY8 multi-pulse sequence to suppress single particle dephasing. The consecutive spin echo pulses spaced by time $\tau$ form a bandpass filter that rejects noise outside of a window with center frequency $f_0 = 1/(2\tau)$ and width $\sim 1/T$, suppressing the Ramsey phase fluctuations caused by noise in $\bm{E}$ and the nonzero differential ac polarizability. The XY8 sequence's time-reversal symmetry and alternation between rotation axes further improve robustness against pulse area error and finite pulse duration. Multiple XY8 sequences are concatenated to achieve longer Ramsey time $T$ without affecting $f_0$.

The efficacy of the dephasing suppression depends on $\tau$, pulse duration $t_p$, and the timing of the sequence, which we optimize with the following procedure:

An ideal XY8 sequence consists of infinitely fast echo pulses. However, if the microwave Rabi frequency is too high, molecular hyperfine states that are not in the spin-1/2 manifold can be coupled off-resonantly by the microwave field. To avoid this effect, we limit the Rabi frequency such that the $\pi$ pulses time is around 10 μs for measurements in $\{|\downarrow\rangle, |\uparrow_1\rangle\}$, and around 16 μs for measurements in $\{|\downarrow\rangle, |\uparrow_2\rangle\}$.

At fixed microwave power, we precisely determine the pulse duration in order to reduce the rotation error of the pulses. To do this, we perform a $\pi/2$ rotation followed immediately by up to eight closely spaced echo pulses. We fine tune the pulse duration $t_p$ until equal population of the two spin states are obtained regardless of the number of echo pulses applied.

The pulse spacing $\tau$ determines the passband frequency of the noise filter. Shorter $\tau$ gives higher passband frequencies and less sensitivity to low frequency electric field noise, which is the dominant noise source in our system. We characterize noise rejection as a function of $\tau$ by measuring the Ramsey phase fluctuations after a fixed $T$ with $\{|\downarrow\rangle, |\uparrow_1\rangle\}$. We vary $\tau$ by changing the number of equally spaced Rabi-$\pi$ pulses inserted between the two Ramsey pulses. We choose the phase of the second Ramsey pulse to maximize the sensitivity of the fraction of $|\uparrow_1\rangle$ to the Ramsey phase. For each $\tau$, we repeat the measurement 10 times to extract the SD of Ramsey phase, as computed from the measured fraction of $|\uparrow_1\rangle$. We use echoes along $\hat{X}$ only for this experiment, which provides the same electric field noise rejection performance as XY8.

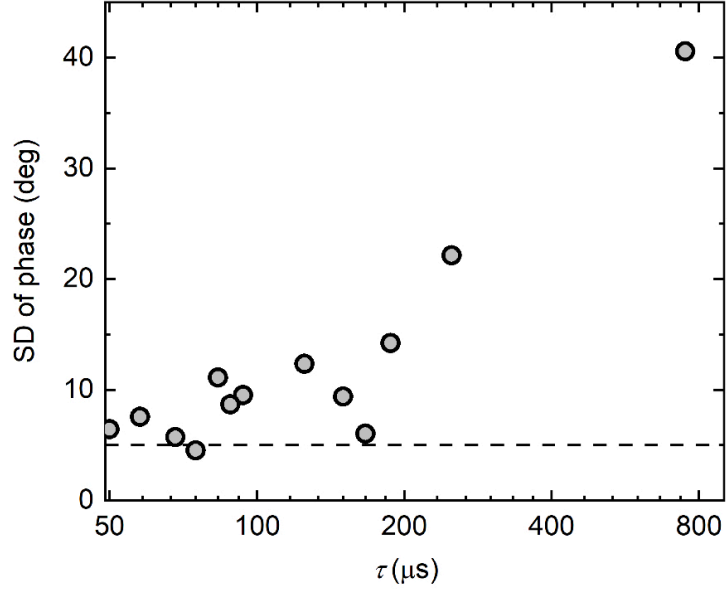

**Extended data figure 2. Noise suppression of the dynamical decoupling sequence.** Data is taken at $|E| = 1.02$ kV/cm and $\alpha = 0°$. We use X-echo only, which provides similar performance to XY8 in terms of rejecting noise in $\Delta$. Each data point is extracted from 10 repetitions. The dashed line is the noise floor of our measurement.

Extended data figure 2 shows the SD of Ramsey phase as a function of $\tau$ for $T = 1.5$ ms. We observe reduction of the phase fluctuations as $\tau$ is reduced. To balance between decoupling efficacy and a high $\tau/t_p$ ratio to minimize system evolution during the pulses, we chose $\tau = 140$ μs for the mean-field frequency shift measurements.

To account for the finite widths of the Ramsey pulses, we adjusted the delay ($\tau/2$ for infinitely short pulses) after the first Ramsey pulse and before the final detection pulse to minimize the sensitivity of the Ramsey phase shift to detuning $\Delta$ at resonance $(\partial\Delta\phi/\partial\Delta)_{\Delta=0}$ and maximize the range of $\Delta$ within which the phase shift is insensitive to $\Delta$ at first order. The amount of the adjustment is determined by simulating the XY8 sequence on a two-level system, and depends on the rotation angle $\theta$, which is accounted for in the measurements.

**Extracting phase shifts for the reversal measurements**

Switching between two spin manifolds requires changing the frequency of the microwave source. We use two procedures to reduce extra phase shift introduced by this process: (1) we measure the phase difference $\Delta\Phi = \Delta\phi(\theta = \pi/4) - \Delta\phi(\theta = 3\pi/4)$ between the Ramsey fringes obtained for $\theta = \pi/4$ and $\theta = 3\pi/4$; (2) For $\theta = 3\pi/4$ measurement, we first prepare the molecules in $|\uparrow_2\rangle$ instead of $|\downarrow\rangle$ and then apply a $\theta = \pi/4$ pulse. This allows us to keep the timings of the sequences for $\theta = \pi/4$ and $\theta = 3\pi/4$ identical.

**Dipolar collisions with transition dipole moment**

In Ref. 22, the dipolar elastic collision rate between spin polarized KRb in 2D was measured. With $n \approx 5.0 \times 10^7$ cm$^{-2}$, $T_0 \approx 250$ nK, $d = 0.2$ D, Ref. 22 reported $\Gamma_0 = 168(48)$ s$^{-1}$. Scaling $\Gamma_0$ as $\Gamma \propto n\sqrt{T_0}\, d^4$ with $n = 1.1(1) \times 10^7$ cm$^{-2}$, $T_0 = 463(9)$ nK, and $d = d_{\downarrow\uparrow}/\sqrt{2}$ (conditions for the grey circles in Fig. 4**a**), we obtain $\Gamma = 95(29)$ s$^{-1}$.

# Acknowledgement

We thank Thomas Bilitewski and Ana Maria Rey for inspirational discussions and critical reading of the manuscript. We acknowledge funding from the DOE Quantum System Accelerator, the National Science Foundation QLCI OMA-2016244, the National Science Foundation Phys-1734006, and the National Institute of Standards and Technology. J. S. H. acknowledges support from the National Research Council postdoctoral fellowship. C.M. acknowledges the NDSEG Graduate Fellowship.

# Author contributions

All authors performed the experiments, analyzed the data, and contributed to interpreting the results and writing the manuscript.

# Data availability

The datasets generated and analyzed during the current study are available from the corresponding authors on reasonable request.

# Competing interests

The authors declare no competing interests.

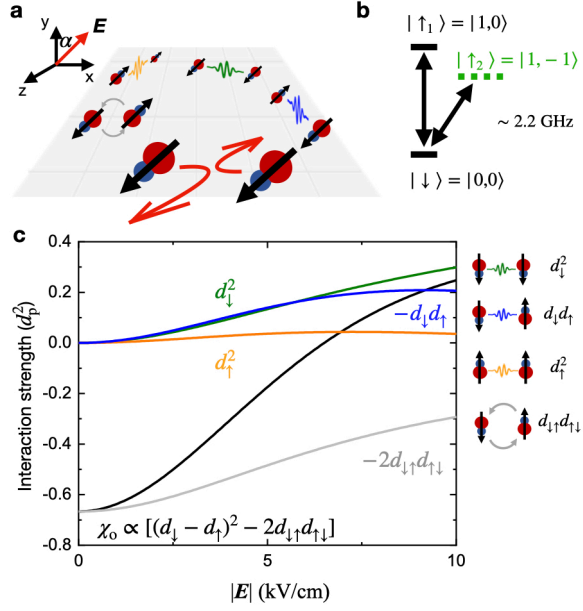

**Figure 1. A 2D itinerant spin system with polar molecules. a.** Molecular spins are free to move within a 2D layer. Their dipolar interactions are tuned by a bias electric field $E$ with configurable magnitude and orientation in the $x$-$y$ plane. Yellow, blue, green, and grey curves illustrate interaction processes detailed in **c**. Red arrows represent dipolar elastic collisions. **b.** Energy diagram for the ground and first excited rotational states in which we encode the spin-1/2 degree of freedom. The transitions are driven by microwaves. **c.** Calculated dipolar interaction strength as a function of $|E|$ for the spin manifold $\{|\downarrow\rangle, |\uparrow_1\rangle\}$. The dipolar coupling strength is given in units of the permanent dipole moment of KRb $d_p = 0.574$ Debye. The molecules interact via their induced dipole moments $d_\downarrow$ and $d_\uparrow$, as well as the transition dipole moment $d_{\downarrow\uparrow}$. The black line shows the field dependence of $\chi_o$.

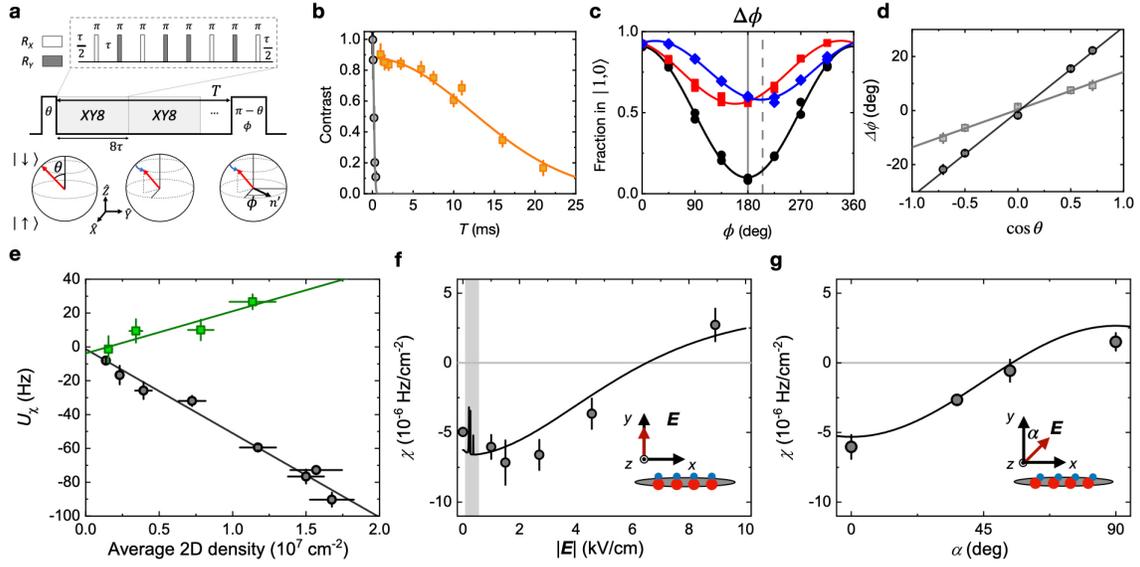

**Figure 2. Dynamical decoupling and tunable dipolar interactions between molecules. a.** The standard Ramsey sequence consists of an initial pulse of area $\theta$, followed by an evolution time $T$, and finally a pulse of area $\pi - \theta$ about an axis $\hat{n} = \hat{X}\cos\phi + \hat{Y}\sin\phi$. We insert one or more XY8 sequences during $T$, denoted as XY8×$M$. One XY8 sequence (zoomed region) consists of 8 Rabi-$\pi$ pulses spaced by time $\tau$, for a total free precession time of $8\tau$ (see Methods). **b.** Decay of Ramsey fringe contrast between $|\downarrow\rangle$ and $|\uparrow_1\rangle$ without dynamical decoupling (grey circles) and with XY8×3 at a density of $n = 0.14(2) \times 10^7$ cm$^{-2}$ (orange squares). We use a low density to reduce interaction effects. Error bars are 1 standard deviation from bootstrapping (see Methods). **c.** Measured Ramsey fringes with dynamical decoupling for an average 2D density $n = 1.4(1) \times 10^7$ cm$^{-2}$ for initial population imbalance $\theta = \pi/4$ (blue diamonds), $\pi/2$ (black circles) and $3\pi/4$ (red squares) for $T = 1.2$ ms at $|E| = 0$. Solid lines are fits to $A\cos[(\pi/180\phi) - \Delta\phi]$ plus an offset, from which we extract the measured shift $\Delta\phi$. **d.** Measured $\Delta\phi$ versus $\cos\theta$ for $n = 1.4(1) \times 10^7$ cm$^{-2}$ (black circles) and $0.65(7) \times 10^7$ cm$^{-2}$ (grey squares). The strength of the mean-field interaction $U_\chi$ is extracted from the slope of the linear fit (solid lines). Error bars are 1 s.e.. **e.** Density dependence of $U_\chi$ for the $\{|\downarrow\rangle, |\uparrow_1\rangle\}$ manifold (grey) and the $\{|\downarrow\rangle, |\uparrow_2\rangle\}$ manifold (green). Black and green solid lines are linear fits to the data. The slopes of the fits are direct measurements of $\chi_1$ and $\chi_2$ respectively. Error bars are 1 s.e. from linear fits. **f.** Dependence of $\chi$ for the $\{|\downarrow\rangle, |\uparrow_1\rangle\}$ manifold on $|E|$ at $\alpha = 0°$. The solid line is a one parameter fit to $A[(d_\downarrow - d_\uparrow)^2 - 2d_{\downarrow\uparrow}d_{\uparrow\downarrow}]$ calculated for KRb at the experimental magnetic field and trapping conditions which includes modifications to the dipole moments from the mixing of hyperfine and rotational states at $|E|$ below ~500 V/cm (shaded area). Grey line indicates zero. Error bars are 1 s.e. from linear fits. **g.** Angular dependence of $\chi$ for the $\{|\downarrow\rangle, |\uparrow_1\rangle\}$ manifold at $|E| = 1.02$ kV/cm. Solid line is a one parameter fit to $-A(3\cos^2\alpha - 1)$. Grey line indicates zero. Error bars are 1 s.e. from linear fits.

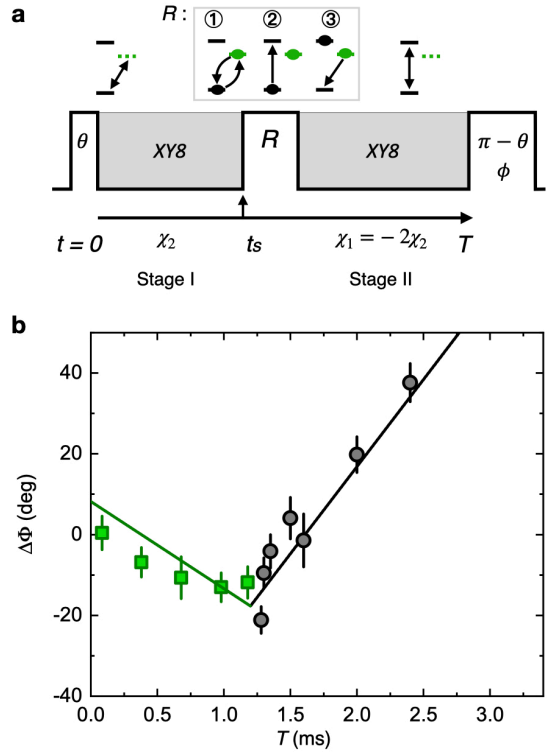

**Figure 3. Reversal of the spin dynamics. a.** Measurement sequence. Molecules are initially prepared and allowed to evolve in a superposition of $|\downarrow\rangle$ and $|\uparrow_2\rangle$. A composite pulse $R$ coherently transfers the population in $|\uparrow_2\rangle$ to $|\uparrow_1\rangle$ while preserving its relative phase to $|\downarrow\rangle$. Molecules then evolve in $\{|\downarrow\rangle, |\uparrow_1\rangle\}$. Inset: Composition of $R$. **b.** Measured time evolution of $\Delta\Phi = \Delta\phi(\theta = \pi/4) - \Delta\phi(\theta = 3\pi/4)$ at $n = 1.1(1) \times 10^7$ cm$^{-2}$. The phase accumulation in stage I (green squares) and stage II (black circles) are plotted over time. For $T < 1.2$ ms, we fix the duration of stage II to be 80 μs and scan the time of stage I. For $T > 1.2$ ms, we fix the duration of stage I to be 1.2 ms and scan the time of stage II. The total time plotted excludes the widths of the microwave pulses. A piecewise linear fit to the time evolution of $\Delta\Phi$ is shown as a solid line. The ratio between the two slopes is constrained to -2. Error bars are 1 s.e..

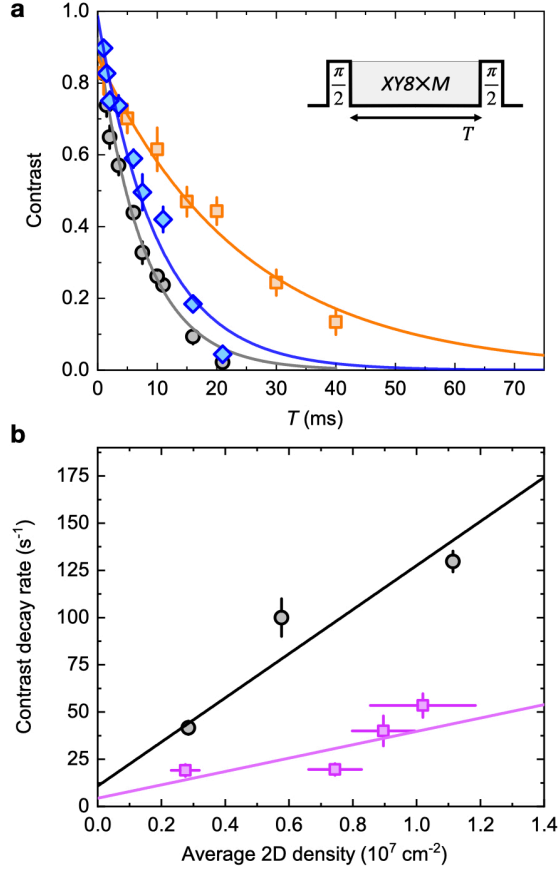

**Figure 4. Dipolar collisional decoherence. a.** Decay of Ramsey contrast between $|\downarrow\rangle$ and $|\uparrow_1\rangle$ for density $n = 1.1(1) \times 10^7$ cm$^{-2}$ (grey circles), $0.57(4) \times 10^7$ cm$^{-2}$ (blue diamonds), $0.28(3) \times 10^7$ cm$^{-2}$ (orange squares) at $|\boldsymbol{E}| = 0$. Decoupling sequences used are XY8×3 for the grey and blue traces and XY8×7 for the orange trace. Error bars are 1 standard deviation from bootstrapping (see Methods). **b.** Contrast decay rate $\Gamma$ as a function of $n$ for $|\boldsymbol{E}| = 0$, $\alpha = 0°$ (grey circles), and $|\boldsymbol{E}| = 1.02$ kV/cm, $\alpha = 36°$ (purple squares). Error bars are 1 s.e. from exponential fits.